\documentclass[11pt]{article}
\usepackage[a4paper, total={6in, 9in}]{geometry}
\usepackage{times}
\usepackage[T1]{fontenc}
\usepackage[latin9]{inputenc}
\PassOptionsToPackage{normalem}{ulem}
\usepackage{ulem}
\usepackage{babel}
\usepackage{graphicx}
\usepackage{xcolor}

\newcommand{\hide}[1]{}
\newcommand{\code}[1]{\textsf{\small #1}}
\newcommand{\comment}[1]{\marginpar{\raggedright\small #1}}

\begin{document}

\title{\vspace*{-8mm}When Are Names Similar Or the Same?\\
Introducing the Code Names Matcher Library%
\setcounter{footnote}{-1}\thanks{This research was supported by the ISRAEL SCIENCE FOUNDATION (grant no.\ 832/18)}}

\author{Moshe Munk ~~~~~~~~~~ Dror G. Feitelson\\
department of Computer Science\\
The Hebrew University of Jerusalem}

\date{}
\maketitle

\vspace{-1ex}
\begin{abstract}
Program code contains functions, variables, and data structures that are represented by names.
To promote human understanding, these names should describe the role and use of the code elements they represent.
But the names given by developers show high variability, reflecting the tastes of each developer, with different words used for the same meaning or the same words used for different meanings.
This makes comparing names hard.
A precise comparison should be based on matching identical words, but also take into account possible variations on the words (including spelling and typing errors), reordering of the words, matching between synonyms, and so on.
To facilitate this we developed a library of comparison functions specifically targeted to comparing names in code.
The different functions calculate the similarity between names in different ways, so a researcher can choose the one appropriate for his specific needs.
All of them share an attempt to reflect human perceptions of similarity, at the possible expense of lexical matching.\\[2mm]
Download: \code{https://pypi.org/project/namecompare/}\\
Code: \code{https://github.com/AutoPurchase/name\_compare}
\end{abstract}

\section{Introduction}

One of the powerful tools for code comprehension research is investigating the names that are given to functions, variables, and data structures.
Names bear witness to what the developer thought about the role of each part of code in the whole program \cite{salviulo14}.
There has therefore been substantial research on names and their meanings \cite{alsuhaibani21,aman21,avidan17,beniamini17,binkley15,butler15b,fakhoury20,gellenbeck91,lawrie07,newman20,schankin18}.

As part of this body of work, it has been shown that different developers often given different names to the same objects \cite{feitelson22}.
And renaming, where a maintainer changes a previously given name, is a common form of refactoring \cite{arnaoudova14,peruma18}.
Additional research has considered how easy it is to remember names \cite{binkley09c,etgar22}, and whether similar names may cause confusion \cite{tashima18,aman21b}.
In all these contexts it is important to be able to compare names to each other.
However, so far this has been done using general string-matching approaches, such as the edit distance.
Even worse, using Python's default Sequence Matcher is susceptible to giving results that depend on the order of the inputs.

To advance the research on variable naming we suggest a library of comparison functions that are specifically designed for comparing names in code.
These functions include features like
focusing on matching words rather than disconnected letters,
dealing with reordering,
and considering semantic similarities.
Importantly, they attempt to promote a metric of similarity that matches human perception, at the possible expense of lexical matching.
Thus they are not as mathematically crisp as the longest common subsequence or edit distance metrics, but may be better tuned for research involving humans and names.

\hide{
confusing names with low edit distance \cite{tashima18}.
similarity of names: \cite{aman21b}.
Aman et al.\ conjecture that similar names may be confusing, especially when dealing with long compound names composed of multiple words.
They therefore study the similarity of 31.8 million name-pairs from 684 Java projects.
Importantly, they distinguish between two types of similarity: string similarity and semantic similarity.
}

\section{Existing Algorithms for String Comparison}

The most common algorithms for comparing two strings (or variable names) are the Longest Common Subsequence algorithm, the Edit Distance calculation, and Python's Sequence Matcher.

\subsection{Longest Common Subsequence}

The Longest Common Subsequence algorithm (LCS) finds the longest sequence of characters that appear in the same order in both strings.
The length of this sequence divided by the length of the shorter (or longer) input string can then be used as a metric for the degree of similarity between the strings.
For example, if the input strings are ``a string is born'' and ``strong is better'', the longest common subsequence is ``strng is br'':
\begin{center}
    \includegraphics[scale=0.8]{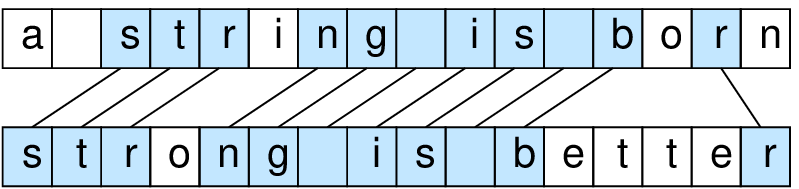}
\end{center}
Thus 11 of 16 characters match, implying that the strings are quite similar to each other.
Note that the letters in the common subsequence need not be consecutive to each other.
When the \emph{are} consecutive we call it a sub\emph{string}.

\subsection{Edit Distance}

The Edit Distance (also known as the Levenshtein Distance \cite{levenshtein66}) counts the minimum number of edit operations required to transform one string into the other.
This has been used in naming research by Tashima et al.\ \cite{tashima18}.
There are several variants depending on which operations are allowed.
The most commonly allowed operations are the insertion, deletion, or substitution of a single character;
a possible fourth is the transposition of adjacent characters \cite{damerau64}.
Using the above two strings as an example again, turning the first into the second requires the following edit operations:
delete 2 characters `a' and ` ';
substitute `i' with `o';
substitute `o' with `e';
add 3 characters `t', `t', and `e';
and delete the character `n':
\begin{center}
    \includegraphics[scale=0.8]{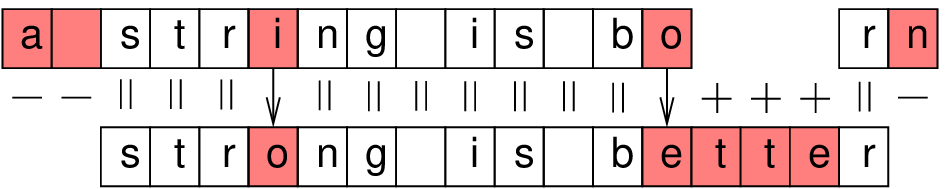}
\end{center}
The total number of character operations which divide the strings is therefore 8.
This can again be turned into a score by dividing by the length of longer input string \cite{aman21b}.
(It is important to normalize relative to the longer string, because the edit distance can be larger than the length of the shorter one but is limited by the length of the longer one.)
Note, however, that this score reflects the divergence between the strings, not their similarity.
To turn it into a similarity score, subtract it from 1.

\subsection{Sequence Matcher}
\label{sect:pyseqmat}

The above algorithms sometimes produce counter-intuitive results.
For example, if we use LCS to compare ``my string has mysterious chars'' with ``mystery match'', the longest common subsequence has 10 letters, but does not include a match of ``mysterious'' with ``mystery''.
This is counter-intuitive to humans, who see this as the most prominent match between these two strings.
\begin{center}
    \includegraphics[scale=0.8]{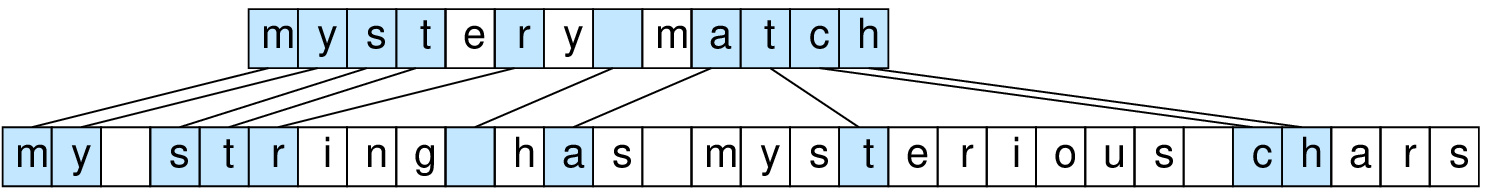}
\end{center}

The Python Sequence Matcher algorithm, which is part of the \code{difflib} library, is intended to better match human intuitions, and has been used in naming research by Etgar et al.\ \cite{etgar22}.
Interestingly, it has also been claimed to be advantageous for comparing sketch-based passwords \cite{amarnadh21}.
It is based on the Ratcliff-Obershelp algorithm, which was designed in the context of educational software to be ``forgiving and understanding of simple typing mistakes, and allow intelligent responses to erroneous input'' \cite{ratcliff88}.
The idea is to prefer matching \emph{contiguous substrings} over matching subsequences of disjoint letters.
Specifically, the algorithm starts by finding the longest contiguous substring of the two input strings, and then continues recursively on both sides of this substring.
Using the above example again, the longest common substring is ``myster'', and the recursive calls will find nothing on the left but a space and ``ch'' on the right, for a total of 9 matching letters.
\begin{center}
    \includegraphics[scale=0.8]{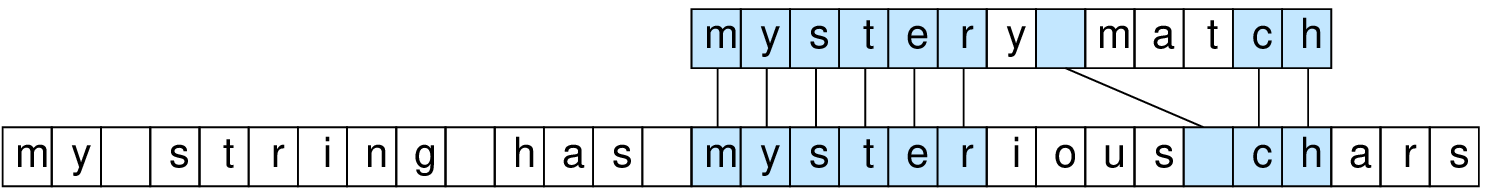}
\end{center}
The final score is the relative length of the resulting subsequence of identical letters, in this case 9/21.5 (the divisor is the average length of the two input strings).

\subsection{Shortcomings of Existing Algorithms}

While all the above algorithms provide reasonable measures for the similarity or difference between strings, they also have shortcomings.
Some of these are especially troublesome in the context of code comprehension studies involving variable names.

One problem, shared by the LCS and Edit Distance algorithms, is the possible fragmentation of the matched subsequences.
For example, if we compare the input string ``stop rampant germs at church'' with the other string ``strange match'', we will find that the shorter string is perfectly matched in the longer one!
But human beings reading these two strings will find this match artificial, because the two strings are composed of words that have no relation to each other.

Another problem is that these algorithms do not deal with changes in order.
As humans we can see that ``first second'' is closely related to ``second first''.
But LCS will just identify the longer repeated part and ignore the other, while calculating the edit distance will double-charge by suggesting a deletion from one place and an insertion in another.

The Python Sequence Matcher algorithm was designed to solve the problem of fragmentation.
However, it too suffers from some drawbacks.
The major one is that it is not symmetric, meaning that comparing string \code{a} with string \code{b} may lead to a different result than comparing \code{b} to \code{a}.
The reason for this unfortunate ``feature'' is that there can be more than one substring with the same maximal length.
The result then depends on which is found first and used as the root of the recursion, and this may depend on the order of the input strings%
\footnote{See \code{\footnotesize https://stackoverflow.com/questions/35517353/how-does-pythons-sequencematcher-work}.}.
For example, if our input strings are ``FirstLightAFire'' and ``LightTheFireFirst'' in this order, we will find that ``First'' is the longest common substring.
But using this as the basis for matching the input strings leads to an empty remainder to the left in the first one, and an empty remainder to the right in the second.
Therefore the recursive calls will not find any additional matches.
However, if we reverse the order of the inputs, ``Light'' becomes the first longest common substring.
In this case the recursive call with the remainders to the right will find an additional match, ``Fire''.

In the context of variable names (and actually also any other text), another issue is concerned with structure.
Humans do not consider names and strings as sequences of letters --- they parse them into words.
So it may be beneficial to actually compare the words in strings and names, rather than just looking for uninterrupted sequences of letters as done by Python's Sequence Matcher.
In addition, acknowledging the role of words opens the door to also consider the use of synonyms, which may be composed of completely different letters, but convey the same meaning.

\section{The New Library}

Our name matching library is based on the following decisions:
\begin{enumerate}
    \item To prefer long \emph{contiguous substrings} over long disjoint subsequences, as in Python's Sequence Matcher.
    As the example above indicates (matching 10 letters with LCS but only 9 with Python's Sequence Matcher), this may lead to \emph{sub-optimal} matchings.
    But it is supposed to be more in line with how humans look at strings.
    \item To be \emph{symmetric}, thus avoiding the main problem with Python's Sequence Matcher.
    This is achieved by basing the match on the longest substring that maximizes the score rather than on the first one that is found.
    \item To support a focus on \emph{words} rather than letters.
    This is achieved by creating two versions of all the functions, one based on comparing letters and the other based on comparing words.
    It gives users the choice of whether to work at the level of letters or words.
    The difference is that respecting word boundaries better reflects the semantics of the names, and avoids situations such as that shown above where ``mystery'' was matched to ``my string''.
    In addition, when comparing words long and short words receive the same weight.
    \item To take possible \emph{reorderings} into consideration.
    This is achieved by creating functions that allow reordering in addition to the original functions which only match in the same order.
    It is important in the case of variable names to acknowledge that variable names like \code{countItems} and \code{itemsCount} are actually the same.
    Users can use the difference between the score of the ordered and unordered functions to see the effect of word reordering.
    \item To support the option of \emph{semantic matching} based on synonyms.
    This is achieved by adding functions that use a dictionary of synonyms when comparing words.
    It is again important for variable name research, for example to recognize that \code{itemsNumber} and \code{itemsCount} are actually more or less the same.
    In addition, words forming singular-plural pairs or abbreviation-expansion pairs are also matched.
\end{enumerate}

\subsection{Provided Functionality}

Based on the above, we wrote a set of functions for finding matches between strings (or variable names).
Almost all of these functions use the concepts of Python's Sequence Matcher --- first finding the longest matching substring, and then matching both sides of the original match.
The difference is that they seek the optimal matches, so the algorithm is necessarily symmetrical.
Specific functions also enable matching between synonyms and singular/plural words, with a configurable ``cost'' for these cases vs.\ identical words.
In addition, we developed another set of functions that also perform ``cross matches'', i.e.\ matches between the left side of the match in one string to the right side of the other string.
For these functions we had to build a new scoring formula, because using the Sequence Matcher formula will give the same score to matches in any order, while clearly the preservation of order should lead to a higher score.

With all these variations, our library provides six main functions for comparing names:
\begin{enumerate}
\item \textbf{Ordered Match}: A function that operates on \emph{letters}, which improves upon Python's Sequence Matcher algorithm by being symmetric and finding the matches that maximize the score.
\item \textbf{Ordered Words Match}: A similar function that operates on \emph{words}.
As a result, each element match is not binary (match or mismatch) as in letters, but can also have intermediate values when words are similar but not identical.
This naturally affects the scoring.
\item \textbf{Ordered Semantic Match}: A function like the previous one, that uses a database of English synonyms and plural words to score semantically similar words.
\item \textbf{Unordered Match}: A function that operates on \emph{letters}, which enables cross matches (from different sides of the previous match).
Note that this is different from just taking the maximum of running the original functions on both orders of the input \cite{etgar22}, because this needs to be done at all levels of the recursion.
\item \textbf{Unordered Words Match}: A function that operates on \emph{words}, that enables cross matches.
\item \textbf{Unordered Semantic Match}: A function like the previous one, that scores on semantically similar words.
\end{enumerate}
Some of these functions accept parameters which define nuances and variations in the matching, for example the threshold value for considering similar words to be a match.
Full details are given below.

All of these functions return detailed information about the found matches, and specifically:
\begin{enumerate}
\item The ``normalized'' version of the input names (with all letters converted to lowercase and with word separators removed), and the list of words extracted from each one (when the function works on words).
\item The locations of all matches that were found, and, when the function works on words, the score of each individual word match.
\item The final score for the similarity between the two names, in the range $[0,1]$.
\end{enumerate}

In addition to the above functions we also have a function that divides a multi-word name to its constituent words, either based on underscores (or some other separator character provided by the user) or on camelCase notation.
This is used to normalize names and allow for valid comparisons that reflect content and not style.

\subsection{The Scoring formula}

\subsubsection{Scoring Rules}

Before committing to a scoring formula, we developed ten desirable rules that the scoring formula should respect.
These rules can then be used as guidelines when we weigh alternative formulas.
The first three are preliminary requirements reflecting basic principles:
\newcounter{myitem}
\begin{enumerate}
    \item \textbf{Style neutrality}:
    the result of comparing names should not depend on style.
    Therefore changes in capitalization should be ignored, and the \_ should be ignored in snake\_case names.

    \item \textbf{Non-optimality}:
    giving a higher score to the biggest match is not a requirement.
    Of course we want to promote good matches, but counting matching letters is not the only and most important yardstick.

    \item \textbf{Normalization}:
    Matches are not absolute but relative to the length of the names.
    If the compared names have different lengths, this should be reflected by lowering the match score.
\setcounter{myitem}{\value{enumi}}
\end{enumerate}
The next two rules are defined by equalities:
\begin{enumerate}
\setcounter{enumi}{\value{myitem}}
    \item \textbf{Symmetry}:
    the result should be the same regardless of the order of the inputs.
    Thus for any matching function
    \[ \mbox{func}( str1, str2 ) = \mbox{func}( str2, str1 ) \]

    \item \textbf{Consistency}:
    different functions should produce the same result in simple cases.
    The definition of ``simple'' cases is cases that are not handled by special versions of the functions.
    As we have special functions to handle multiple words, reordering, and semantic matching, the simple cases are names composed of a single word, with no reordering, and no semantic equivalences.
    For such cases and all matching functions
    \[ \mbox{func1}( str1, str2 ) = \mbox{func2}( str1, str2 ) \]
\setcounter{myitem}{\value{enumi}}
\end{enumerate}
The other five rules are inequalities, which reflect different possible levels of matching:
\begin{enumerate}
\setcounter{enumi}{\value{myitem}}
    \item \textbf{Focus}:
    adding extra baggage reduces the matching score.
    Thus for any matching function
    \[ \mbox{func}( \mbox{"match"}, \mbox{"match"} )
        > \mbox{func}( \mbox{"match"}, \mbox{"match.and.more"} ) \]

    \item \textbf{Continuity}:
    preserving the continuity of adjacent items should always receive a higher score than when they are separated.
    Thus for all functions
    \[ \mbox{func}( \mbox{"begin.end.aaaaaa"}, \mbox{"begin.end.zzzzzz"} ) > \mbox{func}( \mbox{"begin.end.aaaaaa"}, \mbox{"begin.middle.end"} ) \]

    \item \textbf{Order preservation}:
    matching in the same order should always receive a higher score than when words or letters are reordered.
    And for functions that support reordering, matching with reordering should score higher than matching of unrelated strings.
    Thus for all functions
    \[ \mbox{func}( \mbox{"begin.end"}, \mbox{"begin.end"} )
        > \mbox{func}( \mbox{"begin.end"}, \mbox{"end.begin"} ) \]
    and for reordering functions
    \[ \mbox{unordered}( \mbox{"begin.end"}, \mbox{"end.begin"} )
        > \mbox{unordered}( \mbox{"begin.xyz"}, \mbox{"wuv.begin"} ) \]
    
    \item \textbf{Words weight equality}:
    when names are divided into words, long and short words have the same weight.
    Thus letter-matching achieves a higher score than word-matching on long words, and a lower score on short words:
    \[ \mbox{words}( \mbox{"gargantuan\_small"}, \mbox{"gargantuan\_little"} )
        < \mbox{letters}( \mbox{"gargantuan\_small"}, \mbox{"gargantuan\_little"} ) \]
    and
    \[ \mbox{words}( \mbox{"gargantuan\_small"}, \mbox{"humongous\_small"} )
        > \mbox{letters}( \mbox{"gargantuan\_small"}, \mbox{"humongous\_small"} ) \]
    
    \item \textbf{Semantics count}:
    exact matches should always score higher than those based on semantics.
    But for functions that support semantic matching, this should score higher than matching of unrelated words.
    Thus for all functions
    \[ \mbox{func}( \mbox{"little"}, \mbox{"little"} )
        > \mbox{func}( \mbox{"little"}, \mbox{"small"} ) \]
    and for semantic functions
    \[ \mbox{semantic}( \mbox{"little"}, \mbox{"small"} )
        > \mbox{semantic}( \mbox{"little"}, \mbox{"smell"} ) \]
\end{enumerate}

\subsubsection{Crafting a Formula}

It is impossible to reduce all the above rules to practice in a single formula applicable to all the special cases.
We therefore made some compromises.

We start with the scoring formula used in Python's Sequence Matcher, called the \code{ratio}, and use it for scoring the similarity between sequences of letters.
This formula works as follows:
given two strings \code{a} and \code{b}, and letting \code{m} stand for all the matches between them, the similarity score between those two strings is
\[
  ratio = \frac{2|\code{m}|}{|\code{a}|+|\code{b}|}
\]
(where the $|$~$|$ refers to the number of letters in each string).
This score is just the ratio between all the matching letters and the average length of both strings.
It provides normalization, symmetry, and focus from the list of desired rules.

In functions which handle words we depart from the above formula.
On one hand, we want to distinguish close-but-not-identical words.
For example, the score when comparing ``similar'' to itself should be higher than when comparing it to ``similarity''.
But at some point we need to recognize that the words are probably just different, e.g.\ when comparing ``similar'' with ``smaller''.
Our compromise is to use the same ratio formula from above also when comparing words, but to define a minimal required threshold.
If the threshold is not reached the similarity score between the words is set to 0.
Note that for dissimilar words this breaks the requirement for consistency.
The default threshold value is 2/3 (but it can be changed by a parameter).
Using $ratio$ to denote the basic similarity between two words, the word similarity is then
\[
  word\_ratio = \left\{
  \begin{array}{l@{\hspace{10mm}}l}
     ratio  & \mbox{if } ratio \ge 2/3 \\
     0      & \mbox{otherwise}
  \end{array}
  \right.
\]

when the compared names are composed of multiple words, the scores for the different words need to be combined.
This is done using a simple generalization of the basic formula.
Denoting the two inputs by \code{a} and \code{b}, and using \code{m} to denote the set of matching words (that is, words that score above the threshold), the formula is
\[
  multi\_word\_ratio = \frac{\displaystyle 2 \sum_{w \in \code{m}} word\_ratio(w)}
  {||\code{a}||+||\code{b}||}
\]
where $||$~$||$ denotes the number of words in each name.
This gives the same weight to each word.

When using semantic search, we need to decide on the value to assign to two dissimilar words which are found to be synonyms.
We decided to use the threshold value used to identify matching words.
In other words, in the functions that support the matching of synonyms and the such, these matches will be given a score of 2/3 --- at the bottom of the range of scores for matching words.
If the threshold is changed, so is the score for semantic matches.

To score for continuity we consider a string of $n$ elements (letters or words) as if it was composed of $2n-1$ elements.
These elements are the original $n$ elements, and an additional $n\!-\!1$ ``glue'' elements binding neighboring letters or words.
For example, the string ``ab'' will be considered as being composed of ``a'', ``a\&b'', and ``b''.
When comparing it with another string ``cab'' all three will match.
But if the other string is ``acb'', only the ``a'' and ``b'' elements will match, and the ``a\&b'' glue element will not, leading to a lower total score.

The appropriate relative weight of glue elements is debatable.
It could be any arbitrary configurable value, but this would be hard to justify.
We therefore chose to use the highest possible value that respects a preference for longer matching over continuity.
This means that we want $n$ matching elements with no continuity to score higher than $n\!-\!1$ matching elements that are all continuous.
To achieve this, the default score for all the glue elements together should be equal to the score for matching one basic element.
So, when comparing names with $n$ words each, each matching word adds $\frac{1}{n+1}$ to the score, and each matching glue adds $\frac{1}{(n-1)(n+1)}$ to the score.
Importantly, this leads to a perfect score of 1 when the inputs are indeed identical.
If the inputs have different numbers of elements, $n$ is set to the average number, as in the basic matching formula above.

Finally, we need to consider the distinction between ``straight matches'' and ``cross matches''.
As noted in the rules above, it is desirable to give a somewhat lower score to matches that require re-ordering.
Note, however, that scoring for continuity already provides some score for order, because if words match after re-ordering their glue elements will not.
We therefore decided to avoid the overhead of measuring any additional deviations in order, and make do with scoring continuity as detailed above.
The price is that a comparison of ``one\_and\_two\_and\_three\_and\_four'' with ``one\_or\_two\_or\_three\_or\_four'' will achieve the same score (when reordering is allowed) as a comparison with ``four\_or\_three\_or\_two\_or\_one'': in both cases 4 of 7 words are matched, and no glue elements.

\subsection{Algorithmic Effects}
\label{sect:algo-score}

It should be noted that the scoring depends not only on the formula, but also on algorithmic decisions.
We explicitly decided not to strive for the optimal match, meaning the one that would lead to the highest score.
The reason was a desire to better match human intuitions.

This decision led to the selection of the following algorithms for matching.
When matching letters, we start with the longest consecutive substring, and use it as an anchor.
This may lead to a suboptimal score as shown above in Section \ref{sect:pyseqmat}.
When matching words, we likewise search for the longest sequence of matching words.
Again, this can lead to ``missed'' matches for other words, or not matching the longest words, and a reduced score.

For example, consider the matching of ``multi\_multiplayer'' with ``multiplayers\_layer''.
Matching at the letters level finds ``multiplayer'' as the first match, and then does not find any additional matches, for a final score of 0.665.
But with word matching, ``multi'' is matched with ``multiplayers'', and ``multiplayer'' is matched with ``layer'', which together with the glue connecting them leads to a score of 0.734.
Beyond demonstrating possible differences in matching, this example also shows that it is probably impossible to find a single formulation that will be optimal in some sense for all possible cases.

\subsection{Examples}
\newcommand{\m}[1]{{\color{orange}#1}}
\newcommand{\n}[1]{{\color{magenta}#1}}
\newcommand{\s}[1]{{\color{red}#1}}
\newcommand{\g}[1]{{\color{gray}#1}}
\newcommand{\beg}{\begin{center}
\begin{tabular}{|p{0.38\textwidth}|c|p{0.45\textwidth}|}
\hline
}
\newcommand{\fin}{\hline
\end{tabular}\end{center}
}
\newcommand{\f}[1]{\hline
\multicolumn{3}{|l|}{#1:} \\
\hline
}
\newcommand{\e}[4]{\begin{tabular}{@{}l@{}}
#1 \\ #2
\end{tabular} &
#3 &
\parbox{0.44\textwidth}{#4} \\
\hline
}

We start with a simple example, comparing the multi-word strings ``FirstLightAFire'' and\linebreak ``LightTheFireFirst''.
As noted above, the Python Sequence Matcher function gives different results depending of the otrder of the inputs.
Our ordered matcher finds the optimal match regardless of input order.

\beg
\f{\code{difflib\_match\_ratio()}}
\e{\m{First}LightAFire}{LightTheFire\m{First}}{0.312}{matches 5 of 16 letters}
\e{\m{Light}The\m{Fire}First}{First\m{Light}A\m{Fire}}{0.562}{switch inputs, now 9 of 16 letters match}
\f{\code{ordered\_match()}}
\e{First\m{Light}A\m{Fire}}{\m{Light}The\m{Fire}First}{0.557}{finds optimal matching (the slightly lower score is due to partial continuity)}
\fin

When we match words instead of letters, the score is slightly reduced, because matching a long word match does not add more to the score than matching a short word.
And when stop words are ignored, the score is increased because the denominator is smaller.

\beg
\f{\code{ordered\_words\_match()}}
\e{First\m{Light}A\m{Fire}}{\m{Light}The\m{Fire}First}{0.400}{matches 2 of 4 words and no glue}
\f{\code{ordered\_words\_match(ignore\_stop\_words=True)}}
\e{First\m{Light}\g{A}\m{Fire}}{\m{Light}\g{The}\m{Fire}First}{0.625}{matches 2 of 3 words (ignoring ``a'' and ``the'') and 1 glue}
\fin

If we allow reordering, more letters or words can be matched.

\beg
\f{\code{unordered\_match()}}
\e{\m{FirstLight}A\m{Fire}}{\m{Light}The\m{FireFirst}}{0.867}{14 of 16 letters + 11 of 15 glue}
\f{\code{unordered\_words\_match()}}
\e{\m{FirstLight}A\m{Fire}}{\m{Light}The\m{FireFirst}}{0.600}{3 of 4 words + 0 of 3 glue}
\fin

In words matching, each word is a unit.
This is exemplifid by comparing ``multiword\_name'' with ``multiple\_words\_name''.
Using the default threshold of 2/3, only ``name'' matches.
Reducing the threshold to 0.57 allows ``multiword'' to match ``multiple'', but the ``word'' part is left unmatched.
Reducing the threshold even further, to 0.5, allows ``multiword'' to match ``words'', which adds the benefit of matching adjacent words.
Finally, matching based on letters does even better, as it does not respect word boundaries.

\beg
\f{\code{ordered\_words\_match()} [using the default \code{min\_word\_match\_degree=2/3}]}
\e{multiword\_\m{name}}{multiple\_words\_\m{name}}{0.286}{matches 1 word out of 2.5 and no glue}
\f{\code{ordered\_words\_match(min\_word\_match\_degree=0.57)}}
\e{\m{multi}\n{word}\_\m{name}}{\m{multi}\n{ple}\_words\_\m{name}}{0.452}{matches 2 words out of 2.5, with scores of 0.588 and 1, and no glue}
\f{\code{ordered\_words\_match(min\_word\_match\_degree=0.5)}}
\e{\n{multi}\m{word}\_\m{name}}{multiple\_\m{word}\n{s}\_\m{name}}{0.637}{lower threshold allows matching 2 words and glue for a higher score}
\f{\code{ordered\_match()}}
\e{\m{multiword}\_\m{name}}{\m{multi}ple\_\m{word}s\_\m{name}}{0.857}{matches 13 of 15 letters and 10 glues}
\fin

As another example consider the names ``MultiplyDigitExponent'' and ``DigitsPowerMultiplying''.
the Python Sequence Matcher function matches ``multiply'' and two additional letters.
Our ordered match does the same if the minimal match length is set to 1.
With the default minimal required match of 2 it only matches ``multiply''.

\beg
\f{\code{difflib\_match\_ratio()}}
\e{\m{Multiply}D\m{ig}itExponent}{DigitsPower\m{Multiplyi}n\m{g}}{0.465}{10 letters out of 21.5}
\f{\code{ordered\_match(min\_len=1)}}
\e{\m{Multiply}D\m{ig}itExponent}{DigitsPower\m{Multiplyi}n\m{g}}{0.460}{lower score due to partial continuity}
\f{\code{ordered\_match()} [using the default \code{min\_len=2}]}
\e{\m{Multiply}DigitExponent}{DigitsPower\m{Multiply}ing}{0.371}{8 letters of 21.5 and partial continuity}
\fin

When matching words, if we require exact matches then none are found.
But ``digit'' matches ``digits'' with a score of 0.909, and
``multiply'' matches ``multiplying'' with a score of 0.842.
Due to the order constraint, only one of these can be used, and the higher one is selected.

\beg
\f{\code{ordered\_words\_match(min\_word\_match\_degree=1)}}
\e{MultiplyDigitExponent}{DigitsPowerMultiplying}{0.000}{no perfect matches}
\f{\code{ordered\_words\_match()} [using the default \code{min\_word\_match\_degree=2/3}]}
\e{Multiply\m{Digit}Exponent}{\m{Digit}\n{s}PowerMultiplying}{0.227}{one out of 3 words matches with a score of 0.909, and no glue}
\fin

if semantic matching is used, ``exponent'' is found to be related to ``power''.
If we require perfect matching, by setting a matching threshold of 1, this is applied to this semantic match, and there are no additional perfect matches.
With the default matching threshold the semantic match is combined with a regular partial match.

\beg
\f{\code{ordered\_semantic\_match(min\_word\_match\_degree=1)}}
\e{MultiplyDigit\s{Exponent}}{Digits\s{Power}Multiplying}{0.250}{one semantic match considered a perfect match, and no glue}
\f{\code{ordered\_semantic\_match()} [using the default \code{min\_word\_match\_degree=2/3}]}
\e{Multiply\m{Digit}\s{Exponent}}{\m{Digit}\n{s}\s{Power}Multiplying}{0.519}{regular match scoring 0.909, semantic match with default 2/3, and one glue}
\fin

And all the above options can also be combined with reordering.

\beg
\f{\code{unordered\_match(min\_len=1)}}
\e{\m{MultiplyDigit}Ex\m{po}n\m{en}t}{\m{Digit}s\m{Po}w\m{e}r\m{Multiply}i\m{n}g}{0.782}{17 matching letters and 12 glues}
\f{\code{unordered\_match()} [using the default \code{min\_len=2}]}
\e{\m{MultiplyDigit}Ex\m{po}nent}{\m{Digit}s\m{Po}wer\m{Multiply}ing}{0.693}{2 is the default to avoid matching anagrams}
\f{\code{unordered\_words\_match(min\_word\_match\_degree=1)}}
\e{MultiplyDigitExponent}{DigitsPowerMultiplying}{0.000}{no perfect matches}
\f{\code{unordered\_words\_match()} [using the default \code{min\_word\_match\_degree=2/3}]}
\e{\m{MultiplyDigit}Exponent}{\m{Digit}\n{s}Power\m{Multiply}\n{ing}}{0.438}{two matches with scores of 0.909 and 0.842, and no glue matches}
\f{\code{unordered\_semantic\_match()} [using the default \code{min\_word\_match\_degree=2/3}]}
\e{\m{MultiplyDigit}\s{Exponent}}{\m{Digit}\n{s}\s{Power}\m{Multiply}\n{ing}}{0.729}{the above two matches, a semantic match scoring 2/3, and one glue}
\fin

\section{Code Design and implementation}
\newcommand{\class}[1]{\bigskip\noindent\textsf{\textbf{\Large #1}}}
\newcommand{\member}[1]{\textsf{\textbf{\small #1}}}
\newcommand{\method}[1]{\medskip\underline{\textsf{\textbf{#1}}}}

{\setlength{\parindent}{0pt}\setlength{\parskip}{1ex}

\class{class NamesMatcher}

The main class that includes all the functionality is \code{NamesMatcher}.
In its typical usecase it accepts two names (strings) to compare, and a set of parameters that control nuances of the matching.
It then provides a set of functions that return the score for different types of matches, e.g.\ based on letters or words, and with or without reordering.

The constructor for \code{NamesMatcher} is:

\begin{tabbing}
xxxx\=xxxx\=xxxxxxxxxxxxxxxxxxxxxxxxxx\= \+\kill
\code{NamesMatcher(} \+ \\
\code{name\_1=NONE,} \> \# first name \\
\code{name\_2=NONE,} \> \# second name \\
\code{case\_sensitivity=False,} \> \# \code{True} to retain case differences \\
\code{word\_separators='\_~\textbackslash t\textbackslash n',} \> \# to separate words for word matching \\
\code{support\_camel\_case=True,} \> \# to separate words for word matching \\
\code{numbers\_behavior=NUMBERS\_SEPARATE\_WORD,} \# how to treat digits \\
\code{stop\_words\_list=$\left<\right.$\emph{see below}$\left.\right>$} \> \# words to ignore when matching \-\\
\code{)}
\end{tabbing}

The parameters are:

\member{case\_sensitivity}: whether to retain letter case in the comparisons.
The default is to fold all letters to lowercase, so for example \code{cRaZYcaP} is the same as \code{crazyCap}.
If you are analyzing code in a language that distinguishes between upper and lower case you might want to set this to \code{True}.
Also if you are just comparing two strings.
But if you are interested in the \emph{meanings} of names, it is better to ignore case differences.

\member{word\_separators}: letters that signify a word break.
This is only relevant if you are using matching functions that operate on words.
A set of such letters can be provided, and the appearance any one of them in a name indicates a word break.
It is impossible to break only on a sequence of several letters.

\member{support\_camel\_case}:
This too is only relevant if you are using matching functions that operate on words.
The default is to recognize camelCase as an indication for word breaks.

Note: camelCase and word separators may be combined.
Thus ``FileMenu\_saveAsOption'' will be separated into 5 words.

\member{numbers\_behavior}:
how to treat digits that appear in names.
There are three options:
\begin{enumerate}
    \item NUMBERS\_SEPARATE\_WORD = a number is treated as a separate word, namely break words on both sides of the number
    \item NUMBERS\_IGNORE = match only on letters and ignore numbers as if they didn't exist
    \item NUMBERS\_LEAVE = digits are considered to be lowercase letters and conjoined with adjacent letters as one word
\end{enumerate}

\member{stop\_words\_list}: a list of words to ignore when matching words.
The default list includes the words a, are, as, at, be, but, by, for, if, of, on, so, the, there, was, where, were.
This is derived from lists commonly used in information retrieval.
However, some words are intentionally not included, such as ``is'' (which is often used in Boolean variable names like \code{isUpper}) and logical operators (``and'', ``or'', and ``not'').


In place of the constructor, you can also use setters later.
The provided setters are:

\member{set\_name\_1(n)}

\member{set\_name\_2(n)}

\member{set\_names(n\_1, n\_2)}

\member{set\_case\_sensitivity(cs)}

\member{set\_word\_separators(s)}

\member{set\_support\_camel\_case(b)}

\member{set\_numbers\_behavior(nb)}

\member{set\_stop\_words(sw)}

In addition there are getters for all the parameters, and also a couple of special ones for internal representations of the names:

\member{get\_norm\_names()}:
return both names after normalization.
If a name is undefined, \code{None} is returned.
Normalization means three things:
\begin{itemize}
    \item Convert the name to all lowercase (if \code{case\_sensitivity} was not set)
    \item Erase word separators
    \item If \code{numbers\_behavior} was set to \code{NUMBERS\_IGNORE}, erase numbers
\end{itemize}
So the normalized version of ``defaultIs\_FOObar'' is ``defaultisfoobar''.

\member{get\_words()}:
return an array of the words in name.
The procedure for breaking a name into words is:
\begin{enumerate}
    \item Break on the letters identified as word separators, and erase those letters.
    \item Break on both sides of numbers if \code{numbers\_behavior} was set to \code{NUMBERS\_SEPARATE\_WORD}.
    \item Handle camelCase notation, unless \code{support\_camel\_case} was set to \code{False}.
    This is done in two steps:
    \begin{enumerate}
        \item If an uppercase letter is followed by lowercase letters, break before the uppercase one.
        This handles cases like ``camelCase'' $\Rightarrow$ camel+Case, and also ``USAToday'' $\Rightarrow$ USA+Today.
        \item Break on an uppercase letter that appears after a lowercase letter.
        This handles cases like ``theUSA'' $\Rightarrow$ the+USA.
    \end{enumerate}
    If \code{numbers\_behavior} was set to \code{NUMBERS\_LEAVE}, numbers are treated as lowercase letters.
\end{enumerate}
The individual words are normalized as described above.


The main API methods provided by \code{NamesMatcher} are variants of the \code{ratio()} method of Python's Sequence Matcher.
They each perform a different type of matching and compute its score.
Scores are in the range [0,1], with 0 representing no relation whatsoever and 1 representing complete identity.
In addition to the score, these functions also provide full details of the matches that were found.
This is conveyed by an object of class \code{MatchingBlocks}, which is described below.

The API methods are:

\method{ordered\_match()}

This method finds the matches that maximize the ratio between the names.
It is a basic string matching routine, working on individual letters, rather similar to Python's Sequence Matcher.
Logically it starts the matching with the longest consecutive match, and then continues recursively on both sides, thereby maintaining order.
In practice, however, the implementation uses dynamic programming rather than recursion.

The method signature is
\begin{tabbing}
xxxx\=xxxx\=xxxxxxxxxxxxxxxxxxxxxxxxxx\= \+\kill
\code{ordered\_match(} \+\\
\code{min\_len=2,}\\
\code{continuity\_heavy\_weight=False} \-\\
\code{)}\\
\end{tabbing}

The parameters are:

\member{min\_len}: the minimum number of consecutive letters that will be considered as a match.
The default is 2 to avoid the impact of matching individual unrelated letters.

\member{continuity\_heavy\_weight}: the weight given to continuity in the match.
Setting this to \code{True} sets the weight of each continuity element to 1, like the weight of each matching letter, rather than the default where all of them together have a weight of 1.

In addition the method, as all others, is modulated by the parameters of the \code{NamesMatcher} object.
For this method the only relevant parameter is \code{case\_sensitivity}.
To perform simple string matching on the strings as they appear set \code{case\_sensitivity} to \code{True} and \code{min\_len} to 1.

As this method operates on letters and not on words, it does not respect the \code{word\_separators} parameter of \code{NamesMatcher}.
In other words, the word separator letters will be matched like any other letter.

Note: this method uses dynamic programming for its calculation.
As a result its running time is about $m^{2}n^{2}$, where $m$ and $n$ are number of letters in the first and second names, respectively.

\method{unordered\_match()}

This method also finds the matches that maximize the ratio between the names, but it allows cross-matching, that is matches that do no maintain the original order in the two names being matched.

Logically it starts the matching with the longest consecutive match, and then continues recursively to match the concatenations of the leftovers from both sides.
As a result the found matches need not respect the original order.

The implementation, however, is iterative rather than recursive.
Each iteration is composed of two steps.
First, we find the longest consecutive match, and record it.
Then we overwrite the matching letters to block them out.
In each name, the overwriting is done with some letter that does not appear in the other name, to avoid spurious additional matches.
The iterations continue in this manner until no additional matches are found.
Blocking out is used rather than erasing the matching letters because this way the indices of the letters in subsequent matches remain as in the original names.

The method signature and parameters are similar to the previous method:
\begin{tabbing}
xxxx\=xxxx\=xxxxxxxxxxxxxxxxxxxxxxxxxx\= \+\kill
\code{unordered\_match(} \+\\
\code{min\_len=2,}\\
\code{continuity\_heavy\_weight=False} \-\\
\code{)}\\
\end{tabbing}

Note that if \code{min\_len} is set to 1 this method will identify anagrams as perfect matches.

\method{ordered\_words\_match()}

This method finds the word matches that maximize the ratio between the names.
Like the other methods based on words it respects word boundaries, and will not match one word with parts of multiple other words.
    
Logically the method starts by finding the matching sequence of words with the highest total ratio, and continues recursively on both sides, thereby maintaining order.
The implementation, however, uses dynamic programming rather than recursion.
If multiple highest matching sequences are found with exactly the same score, all of them are checked to find the one that leads to the maximal total score.

Note that this method does not guarantee that the found match will include the individual maximal matching word, because some other sequence may lead to a higher total score.
Likewise, it does not guarantee that the final match includes the maximal number of matching words, because other shorted matches may include words with higher individual scores.

The method signature is
\begin{tabbing}
xxxx\=xxxx\=xxxxxxxxxxxxxxxxxxxxxxxxxx\= \+\kill
\code{ordered\_words\_match(} \+\\
\code{min\_word\_match\_degree=2/3,}\\
\code{prefer\_num\_of\_letters=False,}\\
\code{ignore\_stop\_words=False} \-\\
\code{)}\\
\end{tabbing}

The parameters are:

\member{min\_word\_match\_degree}: a float in the range $(0,1]$, which defines the minimum score for two words to be considered as a match.
Setting this to 1 means a perfect match is required.

\member{prefer\_num\_of\_letters}: When there are two or more matches with the same maximal ratio, setting this to \code{True} will cause the best match to be chosen based on the number of letters rather then the number of words.

\member{ignore\_stop\_words}: if \code{True}, ignore any word appearing in the list of stop words and do not count them when scoring the match.
This is only done if there is a perfect match with the word.

Note: this method uses dynamic programming for its calculation.
As a result its running time is about $m^{2}n^{2}$, where $m$ and $n$ are number of words in the first and second names, respectively.

\method{unordered\_words\_match()}

This method also finds the word matches that maximize the ratio between the names, but it allows cross-matching, that is matches that do no maintain the original order in the two names being matched.
Logically the method starts by finding the matching sequence of words with the highest total ratio, and continues recursively to match the concatenations of the leftovers from both sides. As a result the found matches need not respect the original order.
The implementation works by erasing the found match and repeating as long as additional matches are found, as described above for \code{unordered\_match()}.

The method signature and parameters are similar to the previous method:
\begin{tabbing}
xxxx\=xxxx\=xxxxxxxxxxxxxxxxxxxxxxxxxx\= \+\kill
\code{unordered\_words\_match(} \+\\
\code{min\_word\_match\_degree=2/3,}\\
\code{prefer\_num\_of\_letters=False,}\\
\code{ignore\_stop\_words=False} \-\\
\code{)}\\
\end{tabbing}

\method{ordered\_semantic\_match()}

This and the next method are the same as the previous two, except for allowing semantic matches between individual words.
This aims to identify semantically similar names even if different words are used.
The implementation is based on the following considerations regarding what words are considered to be semantically equivalent:
\begin{itemize}
    \item Synonyms are considered to be semantically equivalent and therefore match.
    Synonyms are identified based on a an open thesaurus%
    \footnote{https://raw.githubusercontent.com/zaibacu/thesaurus/master/en\_thesaurus.jsonl}.
    \item Words that differ only in number, that is singular and plural, are considered to be semantically equivalent and therefore match.
    This includes ending with ``s'' or ``es'', and a list of special cases (e.g.\ ``half'' and ``halves'')%
    \footnote{https://github.com/tagucci/pythonrouge/blob/master/pythonrouge/RELEASE-1.5.5/data/WordNet-2.0-Exceptions/noun.exc}.
    \item Abbreviations and their extensions are considered to be semantically equivalent and therefore match.
    This is identified by one word being a prefix of the other, coupled with the requirement that the shorter one be at least 3 letters long.
\end{itemize}

The method signature is
\begin{tabbing}
xxxx\=xxxx\=xxxxxxxxxxxxxxxxxxxxxxxxxx\= \+\kill
\code{ordered\_semantic\_match(} \+\\
\code{min\_word\_match\_degree=2/3,}\\
\code{prefer\_num\_of\_letters=False,}\\
\code{ignore\_stop\_words=False} \-\\
\code{)}\\
\end{tabbing}

The first parameters has an added role:

\member{min\_word\_match\_degree}: a float in the range $(0,1]$, which defines the minimum score for two words to be considered as a match.
This is also used as the score for semantically matching words, when the calculated score is lower.

And the last one may have special significance:

\member{ignore\_stop\_words}: stop words (in, if, of, the, that, etc.) are considered semantically meaningless.
To avoid diluting the score with such meaningless matches, these words can be removed from consideration by setting this parameter to \code{True}.
The list of stop words is set by a parameter to the \code{NamesMatcher} constructor.

\method{unordered\_semantic\_match()}

Same as the previous method, except that reordering of words is allowed.
The reordering is done as in \code{unordered\_words\_match()}.

The method signature and parameters are similar to the previous method:

\begin{tabbing}
xxxx\=xxxx\=xxxxxxxxxxxxxxxxxxxxxxxxxx\= \+\kill
\code{unordered\_semantic\_match(} \+\\
\code{min\_word\_match\_degree=2/3,}\\
\code{prefer\_num\_of\_letters=False,}\\
\code{ignore\_stop\_words=False} \-\\
\code{)}\\
\end{tabbing}

\method{unedit\_match()}

This method is a variant on \code{unordered\_match()}, in which found matches are indeed erased rather than being blocked out.
This enables subsequent matches to span both sides of previous matches, which may be useful in some cases.
Note that this only makes a difference for short matches, where either or both of the matching parts individually are shorter than \code{min\_len}.
    
As the found matches may be discontinuous, this requires a complicated recording of matches locations.

The method signature is
\begin{tabbing}
xxxx\=xxxx\=xxxxxxxxxxxxxxxxxxxxxxxxxx\= \+\kill
\code{unedit\_match(} \+\\
\code{min\_len=2,}\\
\code{continuity\_heavy\_weight=False} \-\\
\code{)}\\
\end{tabbing}


In addition to the above, \code{NamesMatcher} also includes 3 methods implementing traditional string matching algorithms.
These are actually wrappers to implementations in other libraries.
They return the score of the found match, and not a full \code{MatchingBlocks} object like the previous methods.
These methods are:

\method{edit\_distance(enable\_transposition=False)}

An implementation of edit distance calculation, based on insert, delete, and substitute operations.
This has two versions, which come from the \code{strsimpy} library.
The default is to use the \code{Levenshtein()} method.
The alternative, used is \code{enable\_transposition} is set to \code{True}, is to use the \code{damerau()} method.
This adds a fourth basic operation, which is to transpose adjacent letters.

\method{normalized\_edit\_distance(enable\_transposition=False)}

This is an implementation of the edit distance calculation, like the previous method.
The difference is that the calculated edit distance is normalized relative to the length of the longer input.

\method{difflib\_sequence\_matcher()}

This is a wrapper for Python's Sequence Matcher \code{ratio()} function.


\class{class MatchingBlocks}

As noted above, most of the matching methods return all the data about their results in a \code{MatchingBlocks} object.
Such objects contain the following members:

\member{name\_1}: the first name as input to the matching (could be normalized or array of words)

\member{name\_2}: the second name as input to the matching (could be normalized or array of words)

\member{matching\_type}: LETTERS\_MATCH or WORDS\_MATCH

\member{ratio}: the calculated ratio between the two names

\member{matches}: array of matches (\code{OneMatch} objects, describing the locations and lengths of all matches)

\member{cont\_type}: a Boolean indicating if the match is necessarily continuous or not (discontinuous only for \code{unedit\_match()})

\member{continuity\_heavy\_weight}: how glue elements were weighted

\hide{
The following classes are used to store necessary data.
\comment{consider globally replacing ``var'' with ``name''}
\comment{add defined global constants (like default word match threshold?)}

\class{class Var}
\begin{quote}
    \method{Members:}\\
    \member{name}: the original name (string)\\
    \member{words}: list of words that compose the name\\
    \member{norm\_name}: the name after normalization\\
    \member{separator}: a select character that does not appear in the name, for internal use while searching for a match
\end{quote}

\class{class OneMatch}
\begin{quote}
    \method{Members:}\\
    \member{i}: the index of the match in the first name\\
    \member{j}: the index of the match in the second name\\
    \member{k}: the length of the match (in letters or words)\\
    \member{l}: the length of the match in letters when matching words\\
    \member{r}: the ratio (the score for this match)
\end{quote}
}
}

\section{Limitations}

We believe our matching functions provide significant advantages over previous ones, especially for measuring the similarity of names in code.
For example, they avoid the asymmetry of Python's Sequence Matcher.
However, they have their limitations.

In our implementation we preferred breadth over depth:
we wanted to incorporate many ideas, and made do with a basic implementation of each one.
Thus the results may be improved by using better dictionaries or word-splitting procedures \cite{hill14,hucka18}.
In addition, our implementation assumes that it will only be called to compare reasonably short names, and not long texts.
In some of the functions we use dynamic programming or other algorithms that do not scale very well.

One of the perennial problems with splitting is to identify conjoined words like ``multiword'' or ``schoolbus''.
The common approach is to try all possible splittings and compare to a dictionary.
This might be reasonable for 2 words, but not for more.
In our case, if both names include the same conjuncts the problem is moot.
And if one contains the individual base words, or even just some of them, comparing the results of letter-matching with word-matching can provide the necessary information on how to split.

Also, we argued above in Section \ref{sect:algo-score} that matching words should start with the highest single match, even if this comes at the expense of not achieving the highest total score.
A possible alternative may be to check all possible matches, and find the combination that leads to the highest score.
Our choice was based on anecdotal evidence that this leads to matches that seem better.
But it may be that better compromises can be found.

Our scoring function also completely ignores reordering, except as detected by lack of continuity.
And indeed, it seems hard to create a formula that acknowledges both continuity and order preservation.
When names with only two words are considered, continuity and order are actually the same thing, so this would lead to double counting.
This may also happen when there are more words, e.g.\ when the last two are switched.
In other cases we need to divide the extra score between the two attributes in some way.
And if we want to retain the preference for higher scores for more matches, the total attributed to continuity and order is reduced as the number of words grows.
Future work may better resolve these issues.

Finally, our work like the vast majority of work on comparing names places a focus on the matching letters which compose words.
However, the mapping of words to meanings is not one-to-one.
We acknowledge this by providing functions that attempt to take semantics into account by looking for synonyms.
However, it is also important to try to avoid homonyms --- words with the same spelling that mean different things \cite{arnaoudova10}.
A possible approach is to use natural language approaches based on big data to identify words that appear in similar contexts.
However, such approaches suffer from possible confounding of synonyms and antonyms \cite{ali19}.
For example, Table 1 in the groundbreaking paper by Alon et al.\ introducing code2vec includes ``notEqual'' as one of the tokens that appear in the same contexts as ``equal'' \cite{alon19b}.
Dealing with this would require a whole new level of analysis, which we leave to future work.

\bibliographystyle{myabbrv}
\bibliography{abbrv,se,misc}

\end{document}